\documentclass[intlimits,twoside,a4paper]{article}

\usepackage[active]{srcltx}

\usepackage{amsmath,amssymb}
\usepackage{graphicx}
\usepackage{wrapfig}
\usepackage{multirow}
\usepackage{xcolor}
\usepackage[cp1251]{inputenc}

\usepackage{cmpj3}

\issue{2017}{20}{4}{43801}
\doinumber{10.5488/CMP.20.43801}

\title[Fluctuation potential in the modified Poisson-Boltzmann theory]
{An analysis of the fluctuation potential in the modified Poisson-Boltzmann theory for restricted primitive model electrolytes}

\author[E.O. Ulloa-D\'{a}vila, L.B. Bhuiyan]{E.O. Ulloa-D\'{a}vila, L.B. Bhuiyan}

\address{
Laboratory of Theoretical Physics, Department of
Physics, Box 70377, University of Puerto Rico, \\San Juan, Puerto
Rico 00936-8377, USA
}

\date{Received July 14, 2017, in final form August 22, 2017}

\begin{document}

\maketitle

\begin{abstract}

     An approximate analytical solution to the fluctuation potential problem in the
modified Poisson-Boltzmann theory of electrolyte solutions in the restricted
primitive model is presented. The solution is valid for all inter-ionic distances,
including contact values. The fluctuation potential solution is implemented in the
theory to describe the structure of the electrolyte in terms of the radial distribution
functions, and to calculate some aspects of thermodynamics, viz., configurational reduced
energies, and osmotic coefficients. The calculations have been made for symmetric
valence 1:1 systems at the physical parameters of ionic diameter $4.25 \times 10^{-10}$~m, relative
permittivity 78.5, absolute temperature 298~K, and molar concentrations 0.1038,
0.425, 1.00, and 1.968. Radial distribution functions are
compared with the corresponding results from the symmetric Poisson-Boltzmann, and
the conventional and modified Poisson-Boltzmann theories. Comparisons have also been
done for the contact values of the radial distributions, reduced configurational energies,
and osmotic coefficients as functions of electrolyte concentration. Some Monte Carlo simulation
data from the literature are also included in the assessment of the thermodynamic predictions.
Results show a very good agreement with the Monte Carlo results and some improvement for osmotic
coefficients and radial distribution functions contact values relative to these theories.
The reduced energy curve shows excellent agreement with Monte Carlo data for molarities up to 1~mol/dm$^{3}$.

\keywords electrolytes, restricted primitive model, fluctuation potential, modified Poisson-Boltzmann theory

\pacs 82.45.Fk, 61.20.Qg, 82.45.Gj
\end{abstract}

\section{Introduction}

One of the more consistently active areas of research in the statistical
mechanics of fluids over the years has been in the field of Coulomb fluids.
These encompass among others, electrolytes, ionic liquids, molten salts, colloids,
and polyelectrolytes, the practical relevance of which extend from biological systems
to industrial chemical processes. The literature on this is vast and theoretical progress
was limited until the application of liquid state theory \cite{friedman1,friedman2,hill,
mcquarrie,croxton} based on classical statistical mechanics. We would like to cite here
a few of the recent reviews on the subject \cite{levin,henderson,messina}.

 A widely used model used in the development of  formal statistical
mechanical theories of  ionic solutions treats  the solvent as a structureless, continuous
dielectric medium with a relative permittivity~$\epsilon _{\text r}$, and the solute particles as
charged hard spheres of  arbitrary diameters  $d _{i}$  and charges $Z_{s}e$ with
$Z_{s}$ being the valence of species $s$.  This is the so-called primitive model (PM) of ionic solutions.
When the ions are of the same size, it is
called the restricted primitive model (RPM). Computer simulations of the RPM and PM over the years
(see for example, references \cite{card,rasaiah1,valleau1,valleau2,rogde,abramo}) have shown the
usefulness of these models in interpreting experimentally determined structures and thermodynamics
of charged fluid systems. Furthermore, the simulation data have proved to be invaluable in theoretical development.

The statistical mechanics of primitive models in liquid state physics has followed
two broad paths: In the first, the focus is on computing the pair correlation function or the radial distribution function $g_{ij}(r_{i}, r_{j})$  from the inter-molecular pair potential $u_{ij}(r_{i}, r_{j})$  starting from the Ursell-Mayer cluster expansion \cite{friedman1,friedman2,hill}, or the distribution function method
\cite{hill,croxton}. Two main routes are used,viz., the Kirkwood, Bogolubov, Born, Green, Yvon (KBBGY) hierarchies (see for example, reference \cite{croxton}) and the Ornstein-Zernike (OZ) equation
\cite{friedman2,hill,croxton}. The KBBGY hierarchies relate correlation functions for
$n$ and $n+1$ fixed particles, the molecular potential, and a charge parameter $\xi $.
To evaluate the pair correlation function, for example, a closure relation between
the pair correlation function $g_{ij}(r_{i}, r_{j})$ and the next higher order correlation
function, that is, the triplet correlation $g_{ijk}(r_{i}, r_{j}, r_{k})$ must be provided
to break the hierarchy. One such relation is the superposition approximation \cite{hill}.
In the OZ approach, the total correlation between two ions is considered to consist of two parts: the direct correlation function $c_{ij}(r_{i}, r_{j})$  between the two particles, and the indirect correlation
$h_{ij}(r_{i}, r_{j})$, which takes into account the presence of a third particle.
This is clearly shown by the OZ equation (see for example, reference \cite{hill}),
which is often regarded as a definition of the direct correlation function. To solve
the OZ equation, a closure relation between the direct and the total correlation
functions is required. Among the more well known closures are: the Percus-Yevick (PY)~\cite{percus},
the Hyper-netted chain (HNC) \cite{morita}, and the mean spherical approximation (MSA) \cite{blum}.

In the second method, which is our interest in the present work, the focus is on
obtaining the same $g_{ij}(r_{i}, r_{j})$, but through a potential approach to the theory based on the
Poisson's equation. The classical theoretical analysis of electrolyte solutions in this regard
is that of Debye and H\"{u}ckel (DH) \cite{debye}, which is a linearized version of the
corresponding non-linear Poisson-Boltzmann (PB) equation. A key theoretical paper on an
assessment of the inherent approximations in the Poisson-Boltzmann (PB) equation, and
hence in the linearized DH equation is due to Kirkwood \cite{kirkwood}. Kirkwood showed
through a statistical mechanical analysis that the main approximations in the classical
theories are the omission of (i) ionic exclusion volume effects,
and (ii) the fluctuation potential term, which involves the inter-ionic correlations. There have
been many attempts since Kirkwood to improve upon the PB/DH theory notable among which has
been the extensive work done by Outhwaite and co-workers (see for example, references \cite{outh1,outh2,outh3,outh3a,martinez,outh4,molero,outh5,outh6,outh7}),
who within the framework of the PM, have analyzed Kirkwood's methods and obtained
estimates for the fluctuation term. The resulting modified Poisson-Boltzmann (MPB)
approach to ionic solutions is thus based on extending the classical mean electrostatic
potential approach of DH theory by expressing the distribution functions in the
Kirkwood, Bogolubov, Born, Green, Yvon (KBBGY) hierarchies in terms of mean
electrostatic potentials. Essentially, the MPB improves upon the classical PB theory by
incorporating (i) ionic exclusion volume effects, and (ii) inter-ionic correlation effects.
This potential procedure solves for the mean electrostatic potential $\psi (r)$ as opposed to the
integral equations that attempt to solve directly for the radial distribution function
$g_{ij}(r_{i}, r_{j})$. Outhwaite and co-workers \cite{outh3,outh3a,martinez,outh4,molero,outh5,outh6} have
further symmetrized the classical PB theory and the MPB theory so that the Onsager relation,
$g_{ij}(r) = g_{ji}(r)$ is satisfied for a homogeneous fluid. They have also coupled
an exclusion volume term to the symmetrized PB theory, and call it the symmetric
Poisson-Boltzmann (SPB) theory \cite{outh4,molero,outh5}.

	In the MPB theory, the mean electrostatic potential is expressed in terms of the
fluctuation potential $\phi(1,2;3)$ (see for example, reference \cite{outh6}) (3 is the field point,
while there are fixed ions at 1 and 2), which measures deviations from the
superposition principle of Kirkwood \cite{kirkwood}, and, therefore, contains information on
the interionic correlations in the theory. Expressed in terms of the mean potentials, the
fluctuation potential is given by \cite{outh4,outh6}
\begin{equation}\label{eq01}
\psi (1,2;3)=\psi(1,3)+\psi(2,3)+\phi (1,2;3).
\end{equation}
This equation is a statement that the mean potential at field point 3 is the sum of the
direct potentials of particles fixed at 1 and 2, and the correlated potential contribution
at the field point from the simultaneous presence of particles at 1 and 2. As we will see
in the next section, the fluctuation potential can be written in terms of distributions functions
as
\begin{equation}\label{eq02}
\phi (1,2;3)=\frac{1}{4\piup \epsilon _{0}\epsilon _{\text r}}\sum_{s}e_{s}
\int \left[\frac{\rho _{s}(1,2;q)-\rho _{s}(1;q)-\rho _{s}(2;q)}{\textbf{r}_{q}}\right]\rd\textbf{r}_{q}\,,
\end{equation}
where $e_{s}$ is the charge and $\rho _{s}(\{n\};q)$ is the number density of the
$s$-th species of ions at $\textbf{r}_{q}$ with $n$ fixed particles at
${\bf r}_{i}$ ($i = 1, \ldots,n$) with the sum being over all species,
$\epsilon _{0}$ is the vacuum permittivity, and $\epsilon _{\text r}$ the relative permittivity (dielectric constant) of the solvent.

In the simplest language, the fluctuation potential is the inter-ionic correlations
expressed in potential form. The fluctuation potential $\phi (1,2;3)$  obeys a system of
partial, non-linear, differential equations, and for the RPM case, the linearized
version of the equations is given in reference (see for example, reference \cite{outh6,outh7}).
An approximate solution, valid for large inter-ionic separation, under the assumption of
spherical symmetry, was found by Outhwaite \cite{outh2}. One of the main problems in present
MPB theory is the restriction of the fluctuation potential for large inter-ionic separations, where
approximate spherical symmetry is valid. In the present work, an approximate analytical
solution to the fluctuation potential problem is found, that is valid for the whole
range of interionic distances. This solution has an advantage of simplicity that can
provide insight into the eventual fully numerical methods for solving this kind of
problems. The approximate analytical solution for $\phi (1,2;3)$ can serve as a guide to
solving the problem numerically without using the approximations of this research.

	The organization of this paper is as follows. In the following section (section~\ref{sec1}) we start by
giving details of the interaction potentials of the model, a brief introduction to the PB
equation and the MPB theory approach. We then proceed to the main theoretical development
of this work based on the primitive models. In this part, the set of differential equations for the
fluctuation potential in dimensionless form is developed and an approximate solution is
found using ordinary electrostatics.

	In section~\ref{sec2} we utilize solution of the fluctuation potential to present structural
and thermodynamic results for a 1:1 valence RPM electrolyte. We start by showing three-dimensional
plots of the fluctuation potential solution. The plots show the fluctuation potential
at a planar slice passing through the center of the ions for two ionic separations and for the
like and unlike ion cases. A physical interpretation of the results in terms of ionic correlation
energy is presented. To further test the solution's validity, configurational energies, and
osmotic coefficients are calculated and compared to the Monte Carlo (MC) simulation data of
Card and Valleau \cite{card}, and Rasiah, Card, and Valleau \cite{rasaiah1}.

    In section~\ref{sec3} we present some conclusions out of this work and stress the
importance of the approach for future work that may involve a full iterative process
using the solution presented here but without the approximations made.

\section{Model and theory}
\label{sec1}
\subsection{Molecular model}

As indicated in the introduction, the model electrolyte system used
in this study consists of a binary, symmetric valence RPM at room temperature.

    The ion-ion interaction potential in the Hamiltonian is thus
\begin{equation}\label{eq03}
 u_{ij}(r)=\left\{
\begin{array}{ll}
\  \infty, & r<d, \\
\dfrac{e^{2} Z_i Z_j}{4\piup \epsilon _{0}\epsilon _{\text r}r}\,, & r>d,
\end{array}
\right.
\end{equation}
where $Z_{s}$ is the valence of ion species $s$, $e$ is the magnitude of the fundamental
charge, $r$ is the distance between the
centres of two ions of types $i$ and $j$, respectively, and $d$ is the common ionic diameter.
The relative permittivity $\epsilon _{\text r}$ is assumed to be uniform throughout the entire system.

\subsection{Theory}

    The formulation of the SPB and the (traditional) MPB have already appeared elsewhere
in the literature (see for example, references \cite{outh3,outh4,molero,outh5}), and
will not be repeated here. We will restrict ourselves to outlining the main steps leading
to the equations governing the fluctuation potential and their solution.

\begin{figure}[!t]
\centerline{\includegraphics[width=0.25\textwidth]{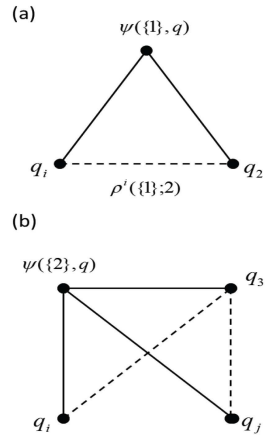}}
\caption{\label{fig1} Diagrammatic representation of the mean electrostatic
potential at field point $q$ due to $n$ fixed charges. Solid lines represents direct
potential, and dotted lines represent potential due to ionic correlation.
(a)~$n = 1$, (b)~$n = 2$.}
\end{figure}

    We begin by formulating the fluctuation potential problem in the restricted
primitive model for a symmetric valence electrolyte, viz., $|Z_{+}|=|Z_{-}|$, consisting
of $N$ ions and satisfying global electroneutrality $\sum _{s}Z_{s}\rho _{s}=0$.
We will closely follow the notations used in reference \cite{outh6}. In the defining
relation for the fluctuation potential in equation (\ref{eq01}), the mean electrostatic potentials
$\psi (1;3)$, $\psi (2;3)$, and $\psi (1,2;3)$ can be formally written as
\begin{equation}\label{eq04}
  \psi (1;3) = \frac{{{e_1}}}{{4\piup \varepsilon _{0}\varepsilon _{\text r} {r_{13}}}} + \frac{1}{{4\piup \varepsilon _{0} \varepsilon _{\text r}}}\sum\limits_\alpha \int {e_\alpha }\frac{\rho _{\alpha }(1,q)}{r_{3q}}\rd q ,
\end{equation}
\begin{equation}\label{eq05}
  \psi (2;3) = \frac{{{e_1}}}{{4\piup \varepsilon _{0}\varepsilon _{\text r} {r_{23}}}} + \frac{1}{{4\piup \varepsilon _{0} \varepsilon _{\text r}}}\sum\limits_\alpha \int {e_\alpha }\frac{\rho _{\alpha }(2,q)}{r_{3q}}\rd q ,
\end{equation}
and,
\begin{equation}\label{eq06}
\psi (1,2;3) = \frac{{{e_1}}}{{4\piup \varepsilon _{0}\varepsilon _{\text r} {r_{13}}}} + \frac{{{e_2}}}{{4\piup \varepsilon _{0}\varepsilon _{\text r} {r_{23}}}} + \frac{1}{{4\piup \varepsilon _{0}\varepsilon _{\text r} }}\sum\limits_\alpha \int e_{\alpha }
\frac{\rho _{\alpha }(1,2;q)}{r_{3q}}\rd q,
\end{equation}
 where $e_{1}$, $e_{2}$ are the charges of the fixed ions at 1 and 2, respectively,
and the sum runs over all the ionic species. Figure~\ref{fig1} shows the mean electrostatic potential at a field point $q$ due to 
1 and 2 fixed ions, respectively.
Subtracting the equations~(\ref{eq04}) and (\ref{eq05}) from equation (\ref{eq06}) leads to
the earlier equation~(\ref{eq02}). The Poisson equations follow
\begin{equation}\label{eq07}
  \nabla ^2\psi (1 ;3) =  - \frac{e_1}{\varepsilon _{0}\varepsilon _{\text r}}\delta (\textbf{r}_{1} - \textbf{r}_{3}) - \frac{1}{\varepsilon _{0}\varepsilon _{\text r}}\sum \limits_\alpha  e_{\alpha }\rho _{\alpha }(1,3),
\end{equation}
\begin{equation}\label{eq08}
  \nabla ^2\psi (2 ;3) =  - \frac{e_2}{\varepsilon _{0}\varepsilon _{\text r}}\delta (\textbf{r}_{2} - \textbf{r}_{3}) - \frac{1}{\varepsilon _{0}\varepsilon _{\text r}}\sum \limits_\alpha  e_{\alpha }\rho _{\alpha }(2,3),
\end{equation}
and,
\begin{equation}\label{eq09}
\nabla ^2\psi (1, 2 ;3) =  - \frac{1}{\varepsilon _{0}\varepsilon _{\text r} }{e_1}\delta (\textbf{r}_{1} - \textbf{r}_{3}) - \frac{1}{\varepsilon _{0}\varepsilon _{\text r}}{e_2}\delta ({\textbf{r}_{2}} - {\textbf{r}_{3}}) - \frac{1}{\varepsilon _{0}\varepsilon _{\text r} }\sum\limits_\alpha  {{e_\alpha }{\rho _\alpha }(1, 2 ;3)}.
\end{equation}
Here, the operator $\nabla $ is understood to operate
on the coordinates of the field point. These equations can also be expressed in terms of
the distribution functions using for example, $g_{1\alpha }(1,q)=\rho _{\alpha }(1,q)/\rho _{\alpha }$,
and so on and so forth, with $\rho _{\alpha }$ being the  mean number density
of ion species $\alpha $.
The distributions can, in turn, be defined in terms of the potentials of mean force $W$, viz., the doublet
\begin{equation}\label{eq10}
g_{ij}(1,2)=\exp[-\beta W_{ij}(1,2)]
\end{equation}
or the triplet
\begin{equation}\label{eq11}
g_{ijk}(1,2,3)=\exp[-\beta W_{ijk}(1,2,3)],
\end{equation}
where $W_{ij}$, $W_{ijk}$ are the pair and triplet potentials of mean force, respectively.
Also, $\beta = 1/(k_{\text B}T)$ with $k_{\text B}$ the Boltzmann constant and $T$ the absolute temperature.
Hence, the conditional distribution,
\begin{equation}\label{eq12}
  g_{ijk}(1,2;3) = {\exp\{ -\beta [W_{ik}(1,3) + W_{jk}(2,3) + {w_{ijk}}(1,2;3)]}\}.
\end{equation}
The term $w_{ijk}$ is the potential of mean force associated with the
departure from linear superposition of the pair potentials. A
hierarchy of such equations can be constructed for higher order correlations. At
the lowest order, the classical PB theory follows upon neglecting $w_{ijk}(1,2;3)$, and to
improve upon the PB, we need a procedure to estimate this term.

In the MPB formulation, the hierarchy is broken at the triplet level by a closure condition
that relates the $w_{ijk}$ with the fluctuation potential $\phi _{ij}$ \cite{outh6}
\begin{equation}\label{eq13}
  {w_{ijk}}(1,2;3) = {e_k}{\phi _{ij}}(1,2;3).
\end{equation}
It is of interest to contrast this MPB closure with the Debye-H\"uckel closure
\begin{equation}\label{eq14}
W_{ij}(1,2) = {e_j}\psi (1;2).
\end{equation}

For the RPM system with a finite ion diameter $d$, the Poisson equations~(\ref{eq07})--(\ref{eq09}) can be expressed in terms of the potentials of mean force as
\begin{align}
  {\nabla ^2}\psi (1;3) &= -
\frac{1}{{{\varepsilon _0}{\varepsilon_{\text r}}}}\sum\limits_s {{e_s}{\rho _s}{\re^{ -
\beta W_{is}(1,3)}}}, \label{eq15}\\
  {\nabla ^2}\psi (2;3) &= -
\frac{1}{{{\varepsilon _0}{\varepsilon
_{\text r}}}}\sum\limits_s {{e_s}{\rho _s}{\re^{ -
\beta W_{js}(2,3)}}},  \label{eq16}\\
  {\nabla ^2}\psi (1,2;3) &= -
\frac{1}{{{\varepsilon _0}{\varepsilon
_{\text r}}}}\sum\limits_s {{e_s}{\rho _s}{\exp\left\{ -
\beta \left[ {W_{is}(1,3) + W_{js}(2,3) + {e_s}\phi (1,2;3)}
\right]\right\}}},\label{eq17}
\end{align}
where the MPB closure (\ref{eq13}) has been used in equation (\ref{eq17}). The equations (\ref{eq15}) and (\ref{eq16}) are exact, for one fixed ion in
position 1 and 2, but equation (\ref{eq17}) incorporates the deviation from the superposition
principle in the form of the fluctuation potential term. To obtain an equation for the
fluctuation potential [equation (\ref{eq01})], we subtract equations (\ref{eq07}) and (\ref{eq08}) from (\ref{eq09}),
\begin{equation}\label{eq18}
  {\nabla ^2}\phi (1,2;3) = -\frac{1}{\varepsilon_{0}\varepsilon_{\text r}}\sum\limits_s{{e_s}{\rho _s}\big[ {g(1,3)g(2,3){\re^{ - \beta{e_s}\phi (1,2;3)}} - g(1,3) - g(2,3)} \big]}.
\end{equation}
Equation (\ref{eq18}) is the base nonlinear equation in the fluctuation potential problem.
The equation also suggests that the charge density source for fluctuation potential is associated
with the charged atmospheres of the triplet and doublet densities.

    To illustrate the geometry of the fluctuation potential problem, one can
expand the summation over species as
\begin{align}\label{eq19}
{\nabla^2}{\phi(1,2;3)} &= - \frac{1}{{{\varepsilon_0}\varepsilon_{\text r}}}{e_{+}}{\rho_{+}} \big[{g(1,{3^{+}})g(2,{3^{+}}){\re^{-\beta {e_{+}}\phi (1,2;3)}} - g(1,{3^{+}}) - g(2,{3^{+}})}\big]\nonumber\\
                        &\quad+ {e_{-}}{\rho _{-}}\big[ {g(1,{3^{-} })g(2,{3^{-} }){\re^{ - \beta {e_{-}}\phi (1,2;3)}} -
g(1,{3^{-} }) - g(2,{3^{-} })} \big],
\end{align}
where a number with a superscript notation with a positive or negative sign represents the presence of the corresponding ion at the referred position in space.

\begin{figure}[!t]
\centerline{\includegraphics[width=0.4\textwidth]{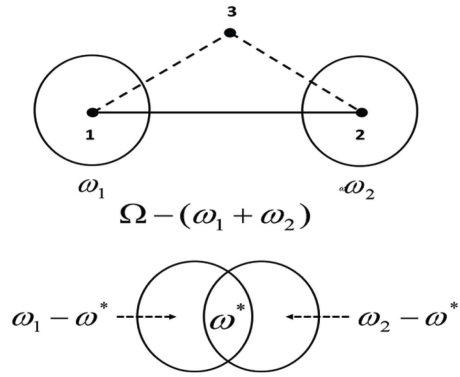}}
\caption{\label{fig2}Geometry of the ionic exclusion volumes within
the restricted primitive model showing the various regions of validity
of the fluctuation potential equation (see text).}
\end{figure}

Figure~\ref{fig2}, represents the geometry of the fluctuation potential
system of equations with $\Omega$ being the total volume of the ionic solution,
$\omega_1$ and $\omega_2$ represent the exclusion volumes of ion 1 and 2,
respectively, $\omega^*$ is the overlap volume, and 3 is the field point.
Region I [$\Omega -(\omega_1 + \omega_2)$]
is the bulk volume defined as the total volume minus the exclusion volumes
of ions 1 and 2. Region II ($\omega_1 - \omega^*$) and III ($\omega_2 - \omega^*$)
are the interior of the exclusion volumes of ion1 and 2 minus the overlap volume.
Region IV is the overlap volume. The nonlinear system of equations
governing the fluctuation potential are then given by the following expressions
\begin{eqnarray}
\text{I}:&  \Omega - ({\omega _1} + {\omega _2})&{\nabla ^2}\phi(1,2;3)= - \frac{1}{{{\varepsilon _0}\varepsilon_{\text r} }}{e_+}{\rho _+}\big[ {g(1,{3^+ })g(2,{3^+ }){\re^{ -
\beta {e_+}\phi (1,2;3)}} - g(1,{3^+ }) - g(2,{3^+ })}
\big]\nonumber\\
&&+ {e_-}{\rho _-}\big[{g(1,{3^-})g(2,{3^-}){\re^{ - \beta {e_-}\phi (1,2;3)}} -g(1,{3^-}) - g(2,{3^-})} \big],\label{eq20}\\
\text{II}: & {\omega _1} - {\omega^*}&{\nabla ^2}\phi (1,2;3) = -\frac{1}{{{\varepsilon_0}\varepsilon_{\text r} }}[{e_+}{\rho _+}g(2,{3^+}) + {e_-}{\rho_-}g(2,{3^-})],\label{eq21}\\
\text{III}:&  {\omega _2} - {\omega ^*}&{\nabla^2}\phi (1,2;3) = - \frac{1}{{{\varepsilon_0}\varepsilon_{\text r} }}[{e_+ }{\rho_+}g(1,{3^+ }) + {e_-}{\rho_-}g(1,{3^-})],\label{eq22}\\
\text{IV}:& {\omega^*}&{\nabla ^2}\phi (1,2;3) =0.\label{eq23}
\end{eqnarray}

At this point it is convenient to work in terms of reduced (dimensionless)
quantities. Here, the relevant ones are the reduced mean electrostatic potential
$\Psi =e\beta \psi$, the reduced fluctuation potential $\Phi =e\beta \phi$, and
$y_{0}=\sqrt{24Z_{+}Z_{-}\eta \Gamma}$. Also, $\eta =(\piup /6)\sum _{s}\rho _{s}d^{3}$
is the volume or packing fraction and $\Gamma = Z_{+}Z_{-}e^{2}/(4\piup \varepsilon _{0}
\varepsilon _{\text r}k_{\text B}Td)$ is the plasma coupling parameter. After expressing the Laplacian
in ionic diameter scale, and imposing global electro-neutrality, we have a set of
dimensionless fluctuation potential equations for the size symmetric case
\begin{eqnarray}
\text{I}:&\Omega - ({\omega _1} +{\omega _2})&-\frac{1}{{y_0^2}}\nabla _d^2\Phi(1,2;3)=  {\frac{{{Z_+}{Z_-}}}{{{Z_-}
- {Z_+}}}}\big[g(1,{3^+ })g(2,{3^+ }){\re^{ - {Z_+}\Phi(1,2;3)}}\nonumber\\
&& - g(1,{3^- })g(2,{3^- }){\re^{ - {Z_-}\Phi (1,2;3)}} - g(1,{3^+ }) - g(2,{3^+ }) + g(1,{3^- }) + g(2,{3^ -})\big],\label{eq24}\\
\text{II}:&\omega _1-\omega ^{*}&-\frac{1}{{y_0^2}}\nabla _d^2\Phi(1,2;3)=  {\frac{{{Z_+}{Z_-}}}{{{Z_-} - {Z_+}}}}[ - g(2,{3^+ }) + g(2,{3^-})],\label{eq25}\\
\text{III}:&\omega _{2}-\omega ^{*}&-\frac{1}{{y_0^2}}\nabla _d^2\Phi(1,2;3)=  {\frac{{{Z_+}{Z_-}}}{{{Z_-} - {Z_+}}}} [ - g(1,{3^+ }) + g(1,{3^-})],\label{eq26}\\
\text{IV}:&{\omega^*}&\nabla _d^2\Phi (1,2;3) = 0.\label{eq27}
\end{eqnarray}
The boundary conditions are that the fluctuation potential and its normal
derivative are continuous across the boundaries.
Denoting the right-hand sides of these equations by ${\rm P}$, we can write them in a general form
\begin{equation}\label{eq28}
\nabla ^2\Phi (1,2;3) = -y_0^2{\rm P}[\Phi ,g(1,3),g(2,3)],
\end{equation}
with a formal solution \cite{griffith,jackson}
\begin{equation}\label{eq29}
\Phi (1,2;3) = \int\limits_\Omega
\ {\frac{{y_0^2}}{{{r_d }}}} {\rm
P}[\Phi ,g(1,3),g(2,3)]\rd{\bf r}_{d}.
\end{equation}
Specifically, we have in the various regions
\begin{eqnarray}
\text{I}:&\Omega - ({\omega _1} + {\omega _2})&- \frac{1}{{y_0^2}}\nabla _d^2\Phi(1,2;3)= {{\rm P}_\text{I}}[\Phi ,g(1,3),g(2,3)] =  {\frac{{{Z_+}{Z_-}}}{{{Z_-} - {Z_+}}}}
\big[g(1,{3^+ })g(2,{3^+ }){\re^{ - {Z_+}\Phi(1,2;3)}}\nonumber\\
&&- g(1,{3^-})g(2,{3^-}){\re^{ - {Z_-}\Phi (1,2;3)}} - g(1,{3^+ }) - g(2,{3^+ }) + g(1,{3^-}) + g(2,{3^ -})\big],\label{eq30}\\
\text{II}:& \omega _1-\omega ^{*}&-\frac{1}{{y_0^2}}\nabla _d^2\Phi(1,2;3) = {{\rm P}_{\text{II}}}[g(1,3),g(2,3)]= {\frac{{{Z_+}{Z_-}}}{{{Z_-} - {Z_+}}}} [ -g(2,{3^+ }) + g(2,{3^ -})],\label{eq31}\\
\text{III}:&\omega _{2}-\omega ^{*}&-\frac{1}{{y_0^2}}\nabla _d^2\Phi(1,2;3) = {{\rm P}_{\text{III}}}[g(1,3),g(2,3)]= {\frac{{{Z_+}{Z_-}}}{{{Z_-} - {Z_+}}}} [ -g(1,{3^+ }) + g(1,{3^-})],\label{eq32}\\
\text{IV}:& {\omega^*}&\nabla_d^2\Phi (1,2;3) ={{\rm P}_{\text{IV}}}, \qquad {{\rm P}_{\text{IV}}} =0.\label{eq33}
\end{eqnarray}

In order to make analytical progress, we approximate the radial distribution functions
$g(1,3)$ and $g(2,3)$, in the various ${\rm P}$'s  appearing in the above equations
by their DH values
\begin{align}\label{eq34}
g(1,3^+ )(=g_{\text{DH}}(1,3^+))   &= \exp \left[ { - {Z_+}{\Psi _{1}^{\text{DH}}}(1,3)} \right],\nonumber\\
g(2,{3^+ })(=g_{\text{DH}}(2,3^+)) &= \exp \left[ { - {Z_+}{\Psi _2^{\text{DH}}}(2,3)}\right],\nonumber\\
g(1,{3^-})(=g_{\text{DH}}(1,3^-))  &= \exp \left[ {- {Z_-}{\Psi _1^{\text{DH}}}(1,3)} \right],\nonumber\\
g(2,{3^-})(=g_{\text{DH}}(2,3^-))  &= \exp \left[ { - {Z_-}{\Psi _2^{\text{DH}}}(2,3)}\right],
\end{align}
where the subscript in $Z$ represents the sign of the charge state of the
ion at the field point 3. Inserting the radial distribution functions~(\ref{eq34}) in the integrals in equation~(\ref{eq29}) will render
the contribution to the fluctuation potential in regions II and III as
ordinary integrals in space.

To obtain an approximation for ${\rm P}$ in the bulk region I, outside
ions 1 and 2, we use the properties of the radial distribution functions in
the various regions, and expand the exponents up to linear terms, leading to
\begin{align}\label{eq35}
{{\rm P}_\text{I}}[\Phi ,g(1,3),g(2,3)] &=  {\frac{{{Z_+}{Z_-}}}{{{Z_-} - {Z_+}}}}[g(1,{3^+ })g(2,{3^+ })(1 - {Z_+}\Phi)- g(1,{3^-})g(2,{3^-})(1 - {Z_-}\Phi )\nonumber\\
& - g(1,{3^+ }) - g(2,{3^+ }) + g(1,{3^-}) + g(2,{3^-})]\nonumber\\
&= {\frac{{{Z_+}{Z_-}}}{{{Z_+} - {Z_-}}}} [{Z_+}g(1,{3^+ })g(2,{3^+ }) - {Z_-}g(1,{3^-})g(2,{3^ -})]\Phi (1,2;3).
\end{align}
For a small fluctuation potential, we neglect the right-hand side of equation~(\ref{eq35}).
For example, for a symmetric valence 1:1 RPM electrolyte, the theme of this work, we have
noted that the DH radial distributions in equation~(\ref{eq34}) are of the order unity for the physical
parameters and the range of concentrations used. If the fluctuation potential is of the order
10$^{-2}$ or less, then the right-hand side of equation~(\ref{eq35}) will be of a similar order and can be
neglected as a first approximation for such a system. Under these approximations, the
fluctuation potential is given by,
\begin{equation}\label{eq36}
\Phi (1,2;3) = \frac{{y_0^2}}{{4\piup }}\left\{{\int\limits_{{\omega _1}}
{\frac{{{F_1}[g(2,{q^+ }),g(2,{q^-})]}}{{\left |
{{{\bf r}_{qd }} - {{\bf r}_{3d
}}} \right |}}\rd{V_q} +
\int\limits_{{\omega _2}}
{\frac{{{F_2}[g(1,{q^+ }),g(1,{q^-})]}}{{\left |
{{{\bf r}_{qd }} - {{\bf r}_{3d
}}} \right |}}\rd{V_q}} } } \right\},
\end{equation}
where
\begin{equation}\label{eq37}
{F_1}[g(2,{q^+ }),g(2,{q^-})] = \frac{{{Z_+}{Z_-}}}{{{Z_-} -
{Z_+}}}[ - g(2,{q^+ }) + g(2,{q^-})],
\end{equation}
and
\begin{equation}\label{eq38}
{F_2}[g(1,{q^+ }),g(1,{q^-})] = \frac{{Z_+}{Z_-}}{{{Z_-} -
{Z_+}}}[ - g(1,{q^+ }) + g(1,{q^-})].
\end{equation}
The integral in equation~(\ref{eq36}) needs to be calculated numerically. This was done
by discretization of space, and will be discussed in the next section.

A useful way of testing the fluctuation potential solution is through
subsequent evaluation of the structure and thermodynamics of the electrolyte
solution. We have utilized the MPB formulation in reference \cite{outh6} to
calculate the pair correlation functions,
\begin{equation}\label{eq39}
g(1,2) = \zeta _{12}\exp \Bigg\{  -
Z_{2}[\Psi (1;2) + \int\limits_0^1 \Phi
(1,2;2)\,\rd\lambda
_{2}]\Bigg\} ,
\end{equation}
where the DH functions~(\ref{eq34}) are used for $\Psi(1;2) $ and an
analytic expression for the Percus-Yevick (PY) radial distribution functions
for hard spheres \cite{mcquarrie} have been used for the excluded volume term,
$\zeta _{12}$. The integral implies charging up of the ion at ${\bf r}_{2}$.

    For the calculation of osmotic coefficients $\phi $ and the reduced configurational
energy $U/(Nk_{\text B}T)$, we use equation~(\ref{eq12}) from reference \cite{burley}, written in
dimensionless reduced variables as,
\begin{equation}\label{eq40}
\phi - 1 = U/3Nk_{\text B}T + 2\eta [{g_\text{A}}(1) + {g_\text{B}}(1)],
\end{equation}
and
\begin{equation}\label{eq41}
U/Nk_{\text B}T =\frac{{y_0^2}}{4}\int\limits_1^\infty\ {[{g_\text{A}}(r')} \ + {g_\text{B}}(r')]r'\rd r',
\end{equation}
where $r'=r/d$ with $g_\text{A}$ and $g_\text{B}$ corresponding to like and unlike ions, respectively, and the argument
1 of $g_{\text A}$ and $g_{\text B}$ in (\ref{eq40}) refers to the contact value.

\section{Results}
\label{sec2}
    All calculations in this work pertain to (1:1) symmetric valence RPM electrolyte
for ions of common diameter $d= 4.25 \times 10^{-10}$~m, in a continuum dielectric
medium of relative permittivity $\epsilon _{\text r}= 78.5$, and  at temperature
$T = 298$~K, which is akin to a water-like solvent at room temperature.  We have
utilized electrolyte concentrations of 0.1038, 0.425, 1.00, and 1.968~mol/dm$^{3}$.
One reason for using these physical parameters is that these have been used earlier
in the literature (see for example, reference \cite{outh7} and for which MC
simulation data exist \cite{card, rasaiah1}. The SPB and the conventional MPB equations
were solved numerically using a quasi-linearization iteration scheme \cite{bellman}.
The procedure has been used with much success in earlier works \cite{martinez,outh4,molero,outh5}
and we refer the reader to these references for further details.

    The fluctuation potential was obtained numerically by solving the integral
in equation~(\ref{eq36}). The radial distribution functions $g_{ij}(r)$ were then calculated
using the fluctuation potential solution in equation~(\ref{eq39}), while the osmotic
coefficient $\phi $, and the reduced configurational energy $-U/(Nk_{\text B}T)$ have been
determined through equations~(\ref{eq40}) and (\ref{eq41}), respectively. In what follows we
will briefly describe the numerical procedure involved before taking up the
discussion of the results.

\subsection{Numerical solution}

\begin{figure}[!b]
\centerline{\includegraphics[width=0.5\textwidth]{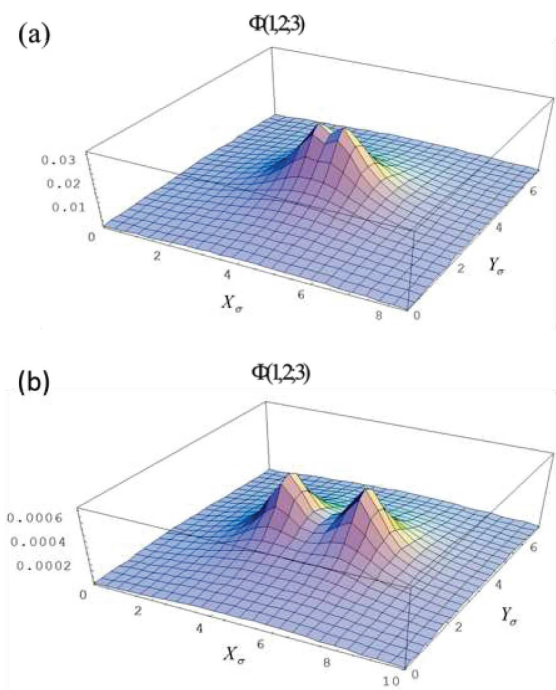}}
\caption{\label{fig3}(Color online) Fluctuation potential $\phi (1,2;3)$ for
$Z_{1} = Z_{2} = +1$ at ionic diameter $d =  4.25 \times 10^{-10}$~m,
dielectric constant $\epsilon _{\text r} = 78.5$, temperature
$T = 298$~K, and electrolyte concentration $c = 1.968$~mol/dm$^{3}$.
Reduced interionic distance $r/d$: (a)~1.5, (b)~3.}
\end{figure}

\begin{figure}[!t]
\centerline{\includegraphics[width=0.5\textwidth]{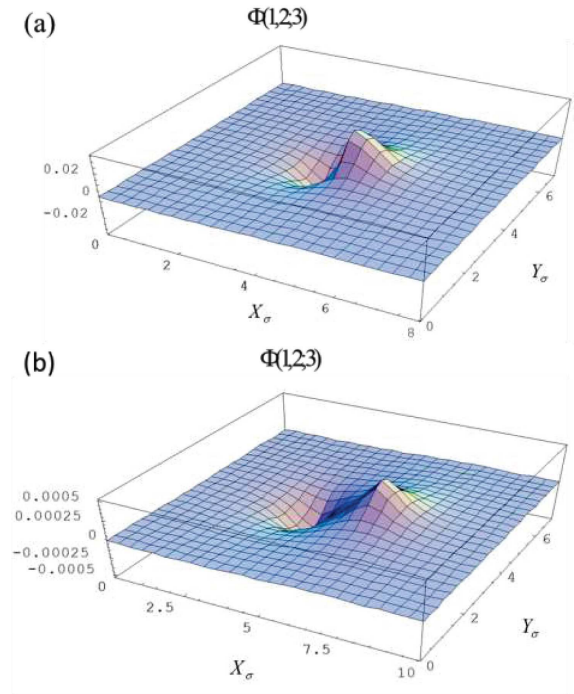}}
\vspace{-1mm}
\caption{\label{fig4}(Color online) Fluctuation $\phi (1,2;3)$ for
$Z_{1} = 1$, $Z_{2} = -1$ at ionic diameter $d =  4.25 \times 10^{-10}$~m,
dielectric constant $\epsilon _{\text r} =$ 78.5, temperature
$T = 298$~K, and electrolyte concentration $c = 1.968$~mol/dm$^{3}$.
Reduced interionic distance $r/d$: (a)~1~(contact), (b)~3.
Note that $r/d= 1$ corresponds to the contact distance.}
\end{figure}

\begin{figure}[!t]
\centerline{\includegraphics[width=0.5\textwidth]{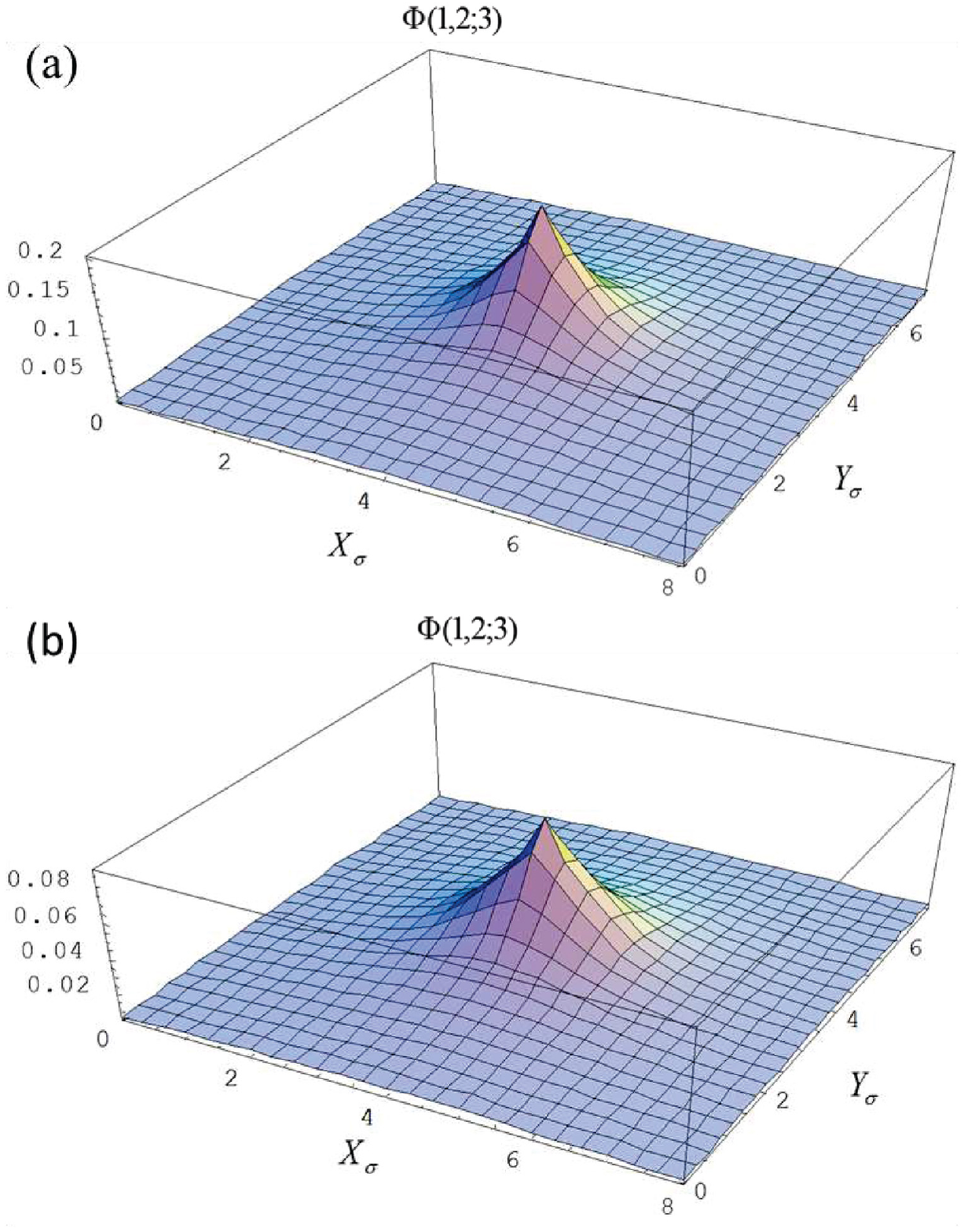}}
\caption{\label{fig5}(Color online) Fluctuation $\phi (1,2;3)$ for
$Z_{1} = Z_{2} = +1$ at ionic diameter $d =  4.25 \times 10^{-10}$~m,
dielectric constant $\epsilon _{\text r} = 78.5$, temperature
$T = 298$~K, and reduced interionic distance $r/d =1$, and
electrolyte concentration: (a) $c = 1.968$~mol/dm$^{3}$, (b) $c = 0.1038$~mol/dm$^{3}$.
Note that $r/d = 1$ corresponds to the contact distance.}
\end{figure}

    The calculation of the fluctuation potential $\Phi(1,2;3)$ was achieved by creating a
Cartesian grid in space with scaled distance of 10\% of the ionic diameter, which in our
dimensionless units is 1, so that in the present context, the grid spacing is 0.1.
This grid was created to represent the physical regions involved in the fluctuation potential
problem as shown in figure~\ref{fig2}. Those regions consist of the spherical regions $\omega _{1}$
and $\omega _{2}$, which correspond to the boundaries of the ions 1 and 2 , and the rest
of the solution region, which is denoted by $\Omega - (\omega _{1} + \omega _{2})$.
The region $(\omega _{1} + \omega _{2})$ is denoted as the ionic excluded volume. The
quantities $F_1$ and $F_2$ [equations~(\ref{eq37}) and (\ref{eq38}), respectively] represent the
charge densities associated with the regions $\omega _{1}$ and $\omega _{2}$
in the integral of equation (\ref{eq36}). The boundary of the rectangular Cartesian grid
representing figure~\ref{fig2} was defined by a parameter $\Lambda $, which represents the distance from
the boundary of the ions to the edge of the grid. This parameter was chosen in such a
way that the fluctuation potential solutions tend to zero at the exterior boundary of the
grid. Usually this parameter was between 3 and 5 ionic diameters for the highest
concentration but was found to a lot larger than at the lower concentrations.
The fluctuation potential solution is an integral over the regions $\omega _{1}$ and $\omega _{2}$.
The summation used to numerically calculate the integral included approximately eight
thousand terms for a point inside regions $\omega _{1}$ and $\omega _{2}$.
To produce the figures~\ref{fig3}--\ref{fig5}, the fluctuation potential was calculated at each point in a
planar slice passing through the centers of $\omega _{1}$ and $\omega _{2}$.
For contact distance between the regions $\omega _{1}$ and $\omega _{2}$, and $\Lambda = 5$,
this planar slice contains approximately ten thousand points. The simplicity of equation~(\ref{eq36})
and the approximation of the $g_{ij}$ in equation~(\ref{eq34}) in terms of the corresponding DH functions
are what makes the calculations fairly tenable.


    The evaluation of the pair correlation functions was performed in a similar
grid as the one used for the three-dimensional figures but now the fluctuation potential
was only required to be calculated at the center of region  $\omega _{2}$ (figure~\ref{fig2}),
and the solution used in equation~(\ref{eq39}), where the Kirkwood charge integral over the
fluctuation potential is calculated.
The calculation of osmotic coefficient and the reduced configurational energy was achieved
using the formulae~(\ref{eq40}) and (\ref{eq41}), respectively.

\subsection{Fluctuation potential}

    We begin this discussion with the analysis of the three-dimensional representations
of the fluctuation potential $\Phi (1,2;3)$ shown in figures~\ref{fig3}--\ref{fig5}. To our best knowledge,
such representation of the fluctuation potential does not presently exist in the literature.
The plots show the fluctuation potential $\Phi(1,2;3)$ obtained from equation~(\ref{eq36}) with
the various $g$'s approximated through equations~(\ref{eq34}). The behaviour pattern of the
fluctuation potential in these figures can be understood in terms of the charge
density associated with the quantities $F_1$ and $F_2$, inside the regions $\omega _{1}$ and $\omega _{2}$.
Figure~\ref{fig3} shows the fluctuation potential for a planar slice passing through the
centers of two positive ions of valence +1 each. The charge density contributed by the
spherical region $\omega _{1}$ due to the positive ion in this region is calculated
using $F_1$ [equation~(\ref{eq37})], which is a function of $g(2,3)$, where the point 3 is
inside region $\omega _{1}$. The positive sign in the fluctuation potential in region $\omega _{1}$
is given by the sign of $-g(2,3^+) + g(2,3^-)$. Since the charge at position 2 is positive,
the second term associated with unlike charges is greater in magnitude than the first term
in $F_1$ causing an overall positive fluctuation potential in region $\omega _{1}$.
The positive sign in region $\omega _{2}$ has similar origins and thus analogous interpretations.

    Figure~\ref{fig4} shows the fluctuation potential for a positive ion (valence $+1$) in
region $\omega _{1}$ and a negative ion (valence $-1$) in region $\omega _{2}$.
In contrast to the situation in figure~\ref{fig3}, in this case the functions $g(1,3)$ and
$g(2,3)$ in $F_1$ and $F_2$ lead to the sign of the fluctuation potential in regions
$\omega _{1}$ and $\omega _{2}$ to be opposite to the signs of the ions 1 and 2,
respectively. To see this, we first look at the fluctuation potential in region
$\omega _{1}$ calculated through $F_1$ with the charge density given by $-g(2,3^+) + g(2,3^-)$.
As the ion in region $\omega _{2}$ is negative, the first term associated with
this unlike charge dominates giving an overall negative sign to the fluctuation potential
in region $\omega _{1}$ where the positive ion is located. On the other hand, the fluctuation
potential in region $\omega _{2}$ is calculated using $F_2$ where the charge density is given
by $-g(1,3^+)+g(1,3^-)$. The second (positive) term here is the larger one in magnitude again
being linked to the unlike charge, and hence the positive sign of the fluctuation potential
in region $\omega _{2}$. So, it can generally be stated that the fluctuation potential for like
ions near the vicinity of these ions is of the same sign as that of the physical ions and is of
the opposite sign for unlike ions. This peculiar behavior is a consequence of the fluctuation
potential in $\omega _{1}$ being related to the $g(2,3)$ centred at 2, and
that the fluctuation potential in region $\omega _{2}$ being related to the $g(1,3)$ centred at the
opposite region $\omega _{1}$. This combined with the relative magnitudes of the $g$'s
in functions $F_1$ and $F_2$ explain the behavior of the polarities in $\Phi (1,2;3)$.

    The  magnitude of the $\Phi (1,2;3)$ that we have noted
in the course of the present calculations, is generally small, especially for large
inter-ionic separations. The reasons for this can again be traced to the dominant
charge density appearing in equation~(\ref{eq36}). For instance, the charge density
in region $\omega _{1}$ is a function of $g(2,3)$ where the field point 3 is
in region $\omega _{1}$ and the point 2 is at the center of region $\omega _{2}$,
and similarly the charge density in region $\omega _{2}$ is a function of $g(1,3)$ where the field
point 3 is in region $\omega _{2}$ and point 1 is at the center of region $\omega _{1}$.
As the inter-ionic separation is increased, the dominant functions in $F_1$ and $F_2$ associated with
the unlike ions decrease, while the $g$'s associated with the like
charges tend to~1. It is clear from equations~(\ref{eq37}) and (\ref{eq38}) that both $F_1$ and $F_2$  tend to zero
at large distances but increase at contact distances, as evident in figure~\ref{fig5}. Significantly,
the fluctuation potential for similar charges is seen to become quite large compared with
that in figure~\ref{fig3}. This suggests that for small separation of the ions, the fluctuation potential term
becomes important in evaluating $g_{ij}$. Figures~\ref{fig3}--\ref{fig5} indeed show that the fluctuation potential is
the largest in the immediate vicinity of ions 1 and 2.

    Another point regarding the fluctuation potential worthy of note is the relationship
between the fluctuation potential and the electrostatic energy of the ions. In figure~\ref{fig3} we
have ions of the same sign, and clearly the fluctuation potential manifests
as an increase in electrostatic energy of the ions since the fluctuation potential is of the same sign
as the ions. For ions of opposite sign as in figure~\ref{fig4}, the sign of the fluctuation potential
is opposite to that of the ion in the vicinity. This leads to a decrease in electrostatic potential energy
leading to attractive inter-ionic correlation in this case. This implies, consistent with what
has been known in the literature, that the sign of the fluctuation potential in the vicinity of ion 1
is mostly due to the cloud of counter ion (from ion 2) and vice versa. It can be seen further from
figures~\ref{fig3} and \ref{fig4}, that the fluctuation potential increases as the separation of the ions decreases,
establishing the importance of having a solution that is valid at short distances. Our results also
show that the fluctuation potential increases with electrolyte concentration.

\subsection{Structure and thermodynamics}

\begin{figure}[!t]
\centerline{\includegraphics[width=0.5\textwidth]{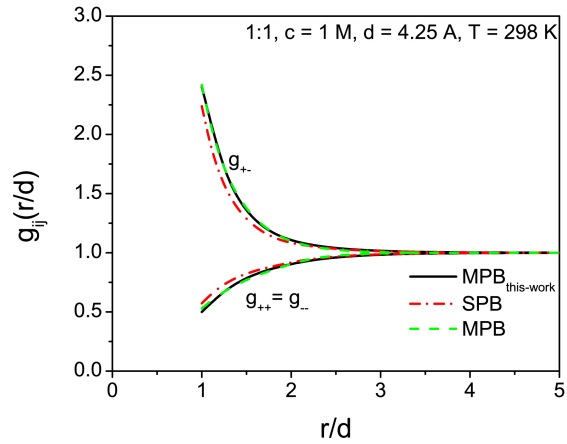}}
\caption{\label{fig6}(Color online) The radial distribution functions $g_{ij}(r)$
 for a 1:1 restricted primitive model electrolyte at ionic diameter
 $d  = 4.25 \times 10^{-10}$~m, dielectric constant $\epsilon _{\text r} = 78.5$,
 temperature $T = 298$~K in the symmetric-Boltzmann theory, the modified
 Poisson Boltzmann theory, and the theory presented in this work. The legend
 as given in the figure.}
\end{figure}

\begin{figure}[!t]
\centerline{\includegraphics[width=0.48\textwidth]{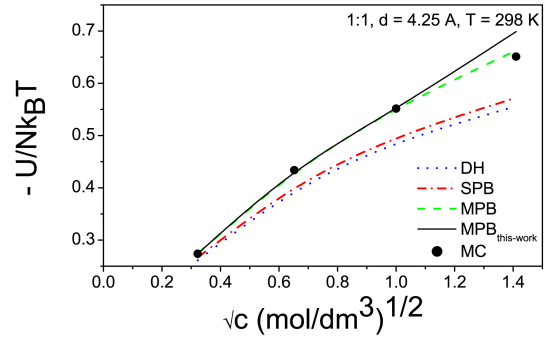}}
\caption{\label{fig7}(Color online) The reduced configurational energy for a 1:1
restricted primitive model electrolyte at ionic diameter $d = 4.25 \times 10^{-10}$~m,
dielectric constant $\epsilon _{\text r} = 78.5$, and temperature $T = 298$~K,
versus the square root of the electrolyte concentration $c$, for the Debye-H\"{u}ckel,
theory, the symmetric Poisson-Boltzmann theory, the modified Poisson-Boltzmann theory,
and the theory presented in this work. Legend as given in the figure.
The Monte Carlo results are from references \cite{card} and \cite{rasaiah1}.}
\end{figure}

\begin{figure}[!t]
\centerline{\includegraphics[width=0.48\textwidth]{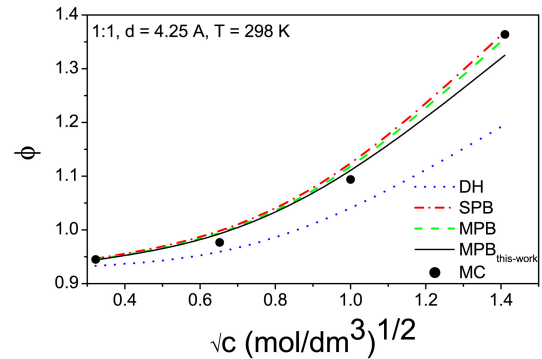}}
\caption{\label{fig8}(Color online) The osmotic coefficient for a 1:1
restricted primitive model electrolyte at ionic diameter $d = 4.25 \times 10^{-10}$~m,
dielectric constant $\epsilon _{\text r} = 78.5$, and temperature $T = 298$~K,
versus the square root of the electrolyte concentration $c$, for the Debye-H\"{u}ckel
theory, the symmetric Poisson-Boltzmann theory, the modified Poisson-Boltzmann theory,
and the theory presented in this work. Legend as given in the figure.
The Monte Carlo results are from references \cite{card} and \cite{rasaiah1}.}
\end{figure}

In figure~\ref{fig6}, we present the radial distribution functions obtained in this work along with
the corresponding curves for the SPB and MPB theories at 1~mol/dm$^{3}$ concentration. It is
clear that the curves are very similar for distances larger than 2 ionic diameters. Importantly,
the present results and the MPB results are almost identical. The contact values for the
radial distribution functions for like ions, from the present theory, are slightly closer
to the MC result \cite{card} than that from the SPB and MPB. This is probably due to a better
treatment of the fluctuation potential in this work. Table~\ref{tab1} shows contact values
$g_{ij}(1)$ and for comparison purposes, the corresponding results from the SPB, the MPB,
and the MC \cite{card, rasaiah1} data are also included. The contact values from the present
theory are consistent with the other theories and show a very good agreement with the MC simulation data.
Tables~\ref{tab2} and \ref{tab3} show reduced configurational energies, and osmotic coefficients from the
Debye-H\"{u}ckel, SPB, MPB, and MC \cite{card, rasaiah1}, and this work. These values
are also presented in a graphic form as in figures~\ref{fig7} and \ref{fig8}, respectively. The reduced
configurational energy curves (figure~\ref{fig7}) show an excellent agreement between the MPB and
this work with the MC curve up to 1~mol/dm$^{3}$ concentration.  At the highest 1.968~mol/dm$^{3}$
concentration, the MPB is a little closer to the MC. Figure~\ref{fig8} shows osmotic
coefficients for the theories and the relevant MC data \cite{card,rasaiah1}. These curves show
a generally very good agreement between the MC results and the theories.

\begin{table}[!t]
\caption{\label{tab1}  Contact values of the radial distribution
functions $g_{ij}(1)$ from different theories. The common diameter
of the ions is $d = 4.25 \times 10^{-10}$~m, the temperature $T= 298$~K, and
the dielectric constant of the electrolyte $\epsilon _{\text r}= 78.5$.
The MC values are from reference \cite{card}.}
\vspace{2ex}
\centerline{\footnotesize{
\begin{tabular}{|c|c|c|c|c|c|c|c|c|c|c|}
\hline\hline
\multirow{2}{*}{$c$ (mol/dm$^{3})$}       &  \multicolumn{5}{c|}{$g_{++}(1)=g_{--}(1)$} &  \multicolumn{5}{c|}{$g_{+-}(1)$}\\
\cline{2-6} \cline{7-11}
  &  DH   &  SPB    & MPB & MPB$_{\text{this-work}}$& MC & DH& SPB& MPB & MPB$_{\text{this-work}}$ & MC\\
\hline\hline
0.1038  & $-$0.158 & 0.321 & 0.311 & 0.302 & 0.319 &  2.16 & 3.19 & 3.30 & 3.33 & 3.25 \\
0.425   &    0.121 & 0.443 & 0.417 & 0.399 & 0.418 &  1.88 & 2.50 & 2.66 & 2.68 & 2.62 \\
1.000   &    0.299 & 0.573 & 0.530 & 0.500 & 0.505 &  1.70 & 2.14 & 2.42 & 2.40 & 2.23 \\
1.968   &    0.433 & 0.752 & 0.686 & 0.633 & 0.706 &  1.57 & 2.20 & 2.40 & 2.31 & 2.38 \\
\hline\hline
\end{tabular}}}
\end{table}

\begin{table}[!b]
\caption{\label{tab2} Reduced configurational
energy $-U/(Nk_{\text B}T)$ from different theories. The common diameter
of the ions is $d = 4.25 \times 10^{-10}$~m, the temperature $T= 298$~K,
and the dielectric constant of the electrolyte $\epsilon _{\text r}= 78.5$.
The MC values are from reference \cite{rasaiah1}.}
\vspace{2ex} \centerline{
\begin{tabular}{|c|c|c|c|c|c|}
\hline\hline
$c$ (mol/dm$^{3\strut})$  &  DH  &  SPB    & MPB & MPB$_{\text{this-work}}$ & MC \\
\hline\hline
0.1038  &    0.261 & 0.267 & 0.274 & 0.274 & 0.274 \\
0.425   &    0.400 & 0.407 & 0.436 & 0.439 & 0.434 \\
1.000   &    0.490 & 0.500 & 0.555 & 0.550 & 0.552 \\
1.968   &    0.556 & 0.572 & 0.663 & 0.699 & 0.651 \\
\hline\hline
\end{tabular}}
\end{table}

\begin{table}[!t]
\caption{\label{tab3}  Osmotic coefficient $\phi $
from different theories. The common diameter of the ions is $d = 4.25 \times 10^{-10}$~m,
the temperature $T= 298$~K, and the dielectric constant of the electrolyte
$\epsilon _{\text r}= 78.5$. The MC values are from reference \cite{card}.}
\vspace{2ex} \centerline{
\begin{tabular}{|c|c|c|c|c|}
\hline\hline
$c$ (mol/dm$^{3\strut})$  &   SPB   & MPB & MPB$_{\text{this-work}}$ & MC \\
\hline\hline
0.1038  &    0.946 & 0.945 & 0.944 & 0.945 \\
0.425   &    0.985 & 0.981 & 0.980 & 0.977 \\
1.000   &    1.11  & 1.10  & 1.10  & 1.094 \\
1.968   &    1.37  & 1.37  & 1.33  & 1.364 \\
\hline\hline
\end{tabular}}
\end{table}


\section{Conclusions}
\label{sec3}
In this study we have made an analysis of the fluctuation potential
in the modified Poisson-Boltzmann theory of bulk electrolyte solutions. An approximate
analytical solution of the fluctuation potential equation was obtained for symmetric
valence 1:1 electrolytes in the RPM. This solution was later utilized to obtain
structural and thermodynamic descriptions of the electrolyte in terms of ion-ion
radial distribution functions, reduced excess energy, and the osmotic coefficients,
respectively.

The fluctuation potential is a central ingredient in a potential approach
to the theory (of charged fluids) such as the modified Poisson-Boltzmann theory. The
fluctuation potential solution developed in this work, albeit with approximations to
make analytical progress and for symmetric 1:1 valence systems, is a preliminary attempt
to assess the implications of such a solution. In such cases, due to the linearization of the
fluctuation potential in the bulk region I [equation~(\ref{eq35})]
and the small magnitude of $\Phi (1,2;3)$, the P function in bulk region I can be
taken to be zero, thus neglecting charge density for that region. A less approximate
 and nearly full treatment could be achieved by solving for the fluctuation
potential in region I using equation~(\ref{eq29}) with P$_{\text I}$ being given by equation~(\ref{eq35}) in conjunction with equation~(\ref{eq34}).  An intermediate
procedure (between the above two situations) to obtain a better, viable, and still feasible
approximation for $\Phi (1,2;3)$ in region I would be to solve equation~(\ref{eq30})
[with P$_{\text I}$ given by equation~(\ref{eq35})] by writing it in the form
\begin{equation}\label{eq42}
\nabla ^{2}\Phi (1,2;3)=\mathcal{C}\Phi (1,2;3),
\end{equation}
where the quantity $\mathcal{C}$ contains the valencies $Z_{+}$, $Z_{-}$, and has spatial dependence
through $g(1,3)$ and $g(2,3)$. Thus, although $\mathcal{C}$ is not a constant per se,
it can be assumed to be approximately constant for the purposes of solution to
equation~(\ref{eq42}). An approximate analytic form of $\Phi (1,2;3)$ in region I,
whose value is not necessarily zero, would then be available. Equation~(\ref{eq42})
has some parallels to a similar equation for the fluctuation potential in the
MPB formalism in the planar electric double layer \cite{outhbari}.
Such a procedure will be useful for higher and multivalent electrolytes when
the magnitude of the fluctuation potential in region I is likely to be significant and hence
P$_{\text I}$ can no longer be neglected. This will be a focus of our future work.

The MPB description of the electric double layer phenomenon is an area where
the present techniques might have some significance since the fluctuation potential plays
an equally important role in the theoretical framework for the inhomogeneous fluid at the
interface. In the MPB approach to the double layer theory in planar \cite{outhbari,bariouth},
cylindrical \cite{outh11,bhuiyan1,bhuiyan2},
and spherical \cite{outh22,bhuiyan3} symmetries, the form of the corresponding fluctuation potential used
is rather approximate and generally suffers from similar defects as those vis-a-vis the
traditional MPB theory for the bulk. The statistical mechanical methods used in this paper
are quite general and can be extended and adapted to interfacial double layer geometry where
an analogous fluctuation potential analysis might prove useful.

 Another area of possible relevance for this study is in the theoretical
analysis of charged fluid systems with a variable dielectric constant (relative permittivity).
The topic has attracted a lot of recent research attention (see for example,
references \cite{guerrera,naji,wang}) and has been shown to be relevant for important technological systems,
viz., super-capacitors \cite{loth,bhuiyan4}. In the electric double layer, the MPB has been found to be
capable of dealing with systems having an inhomogeneous dielectric constant \cite{outhbari,bariouth}.
Very recently, the MPB was applied to a double layer system with three different
dielectric constants \cite{outh33}, although the associated fluctuation potential problem could
only be solved for point ions. Thus, again a fluctuation analysis in such situations
along the lines of the present work could be valuable.

    The three-dimensional plots of the fluctuation potential give a valuable insight
into the correlations between ions. Furthermore, the present structural and thermodynamic
results point in the right direction and are indicative of the potential usefulness of a
full solution of the fluctuation potential. The radial distribution functions, especially
at contact distances between the ions, the reduced excess energy, and osmotic
coefficients show an expected improvement over that from the PB (or SPB), and overall,
tend to be in a very good agreement with the predictions from the traditional MPB theory
and Monte Carlo simulations.

The fluctuation potential problem is a challenging one. A complete solution
of the fluctuation potential equation, valid for a general case and for asymmetry in
ionic size and/or valence will involve a numerical
solution comprising an iterative algorithm. Our solution here might prove useful in
such an involved procedure. Such a project is contemplated in the near future.

\section*{Acknowledgements}

   We are grateful to Professor C.W. Outhwaite of the Department of Applied mathematics,
University of Sheffield, UK, for a critical reading of the manuscript and encouragement.
EOUD would like to thank Dr.~Angel Gonzalez Lizardo, Director of the Plasma Laboratory at
the Polytechnic University of Puerto Rico, for his valuable suggestions and his help with LaTeX.

\ukrainianpart
\title{Аналіз флуктуаційного потенціалу в модифікованій теорії Пуасона-Больцмана обмеженої примітивної моделі електролітів}

\author{E.O. Уллоа-Давіля, Л.Б. Буян}

\address{Лабораторiя теоретичної фiзики, вiддiл фiзики, А/с 70377, Унiверситет Пуерто-Рiко,\\ Сан Хуан, Пуерто-Рiко, США
}

\makeukrtitle

\begin{abstract}
Представлено наближений аналітичний розв'язок проблеми флуктуаційного потенціалу в модифікованій теорії Пуасона-Больцмана для обмеженої примітивної моделі електролітів.  Цей розв'язок є дійсним для всіх міжіонних відстаней,
включаючи контактні значення. Розв'язок для  флуктуаційного потенціалу імплементовано у дану теорію з метою опису структури електроліта в термінах радіальних функцій розподілу, а також з метою обчислення деяких аспектів термодинаміки, а саме, конфігураційної редукованої енергії та осмотичних
коефіцієнтів. Обчислення проведено для систем із симетричною валентністю  1:1 при фізичних параметрах іонного діаметру  $4.25 \times 10^{-10}$~м, 
при відносній проникності  78.5, при абсолютній температурі 298~K, і при молярних концентраціях 0.1038,
0.425, 1.00 і 1.968. 
Радіальні функції розподілу порівнюються з відповідними результатами симетричної теорії Пуасона-Больцмана та стандартної і модифікованої теорій 
Пуасона-Больцмана. Проведено порівняння контактних значень радіальних розподілів, редукованих конфігураційних енергій і осмотичних коефіцієнтів як функцій концентрації електроліта. Деякі дані Монте Карло симуляцій з літератури включено в оцінювання термодинамічних передбачень. Результати
показують дуже добре узгодження з результатами Монте Карло та деяке покращення для осмотичних коефіцієнтів та контактних значень радіальних 
функцій розподілу стосовно вищезгаданих  теорій. Крива редукованої енергії показує чудове узгодження з даними Монте Карло для молярностей аж до  1~моль/дм$^{3}$.

\keywords електроліти, обмежена примітивна модель, флуктуаційний потенціал, модифікована теорія Пуасона-Больцмана

\end{abstract}

\end{document}